\documentstyle[twocolumn,aps,prl,epsf,amssymb]{revtex}

\begin{document} 
\wideabs{    
\draft
\title{Equilibrium and Kinetics: Water Confined in Carbon Nanotube
 as 1D Lattice Gas}
\author{ Xin Zhou, Cheng-Quan Li and Mitsumasa Iwamoto}
\address{Department of Physical Electronics, Tokyo Institute of 
Technology, O-okayama 2-12-1, Meguro-ku, Tokyo 152-8552, Japan }
\date{\today}

\maketitle

\begin{abstract}
A simple 1D lattice gas 
model is presented, which very well describes the equilibrium 
and kinetic behaviors of water confined in a thin carbon nanotube found 
in an atomistic molecular dynamics(MD) 
simulation {[} Nature {\bf 414}, 188 (2001) {]}. The model
parameters are corresponding to various physical interactions and can be 
calculated or estimated in statistic mechanics. 
The roles of every interaction in the water filling, emptying and 
transporting processes are clearly understood.
Our results indicate that the physical picture of the single-file kinetics 
is very simple. 
\end{abstract}
}

 
 Transport of molecules through pores on nanometer scale is important for
 biology. Carbon nanotubes have been widely studied as pores filled with
 gases~\cite{Calbi2001,Skoulidas2002} and 
 liquids~\cite{Dujardin1994,Koga2001}. 
 Recently, using molecular dynamics(MD) simulation, Hummer and 
 coworkers~\cite{Hummer2001} studied the filling and transporting of water 
 molecules in a very thin carbon nanotube which is a $(6, 6)$ tube with 
 two open ends, $4$ {\rm \AA} radius and $5.5$ unit 
 cells (about $13.5$ {\rm \AA} in length).
 The interaction between carbon atoms and water molecules was expressed by 
 Lennard-Jones(LJ) potential. They found that the LJ 
 parameters ($\epsilon$ and $\sigma$) can sensitively affect the 
 filling of water molecules. In their work, two group LJ parameters were used:
 (1) unmodified parameters: $\epsilon = 0.114$ {\rm kcal/mol}, 
 $\sigma = 3.2752$ {\rm \AA}; (2) modified parameters: 
 $\epsilon = 0.065$ {\rm kcal/mol}, $\sigma = 3.4138$ {\rm \AA}.
 For the former, it was found that the water molecules completely filled
  the nanotube (the average occupancy number $n = 5$, the 
 corresponding density is about $5$ times of that of bulk water, where 
 the radius of the nanotube pore is about $0.8$ {\rm \AA}). For the latter, 
 water molecules fluctuated between the empty ($n=0$) and filled ($n=5$) 
 states, and it was a two-state like behavior, the lifetimes of the two 
 states follow exponential distributions. The high density filling 
  seems be surprising in the unmodified system due to the hydrophobic 
  property of carbon nanotube, though 
 the sharper distribution of the water molecule binding energy inside the 
 nanotube was found in the MD simulation~\cite{Hummer2001}. The sensitive 
 effect of LJ potential in the filling of water, the two-state like kinetics, 
 and other remarkable MD results need to more clearly be understood 
 in physics.
   
 Here, based on a strict statistic mechanics treatment, we present a
 one-dimensional(1D) lattice gas(LG) model of water molecules confined 
 in a thin tube with
 a finite length and two open ends. All parameters of this model have explicit
 physical sources, can be calculated from interactions. They are: (1) the local
 excess chemical potential $\Delta {\mu}$ of water in the tube; (2) the free 
 energy contribution ${\cal E}_{c}$
 from LJ potential of carbon atoms; (3) the free energy ${\cal E}_{H}$ 
 from hydrogen bonds between water molecules inside the tube (the rotational 
 free degree of the molecules can be included); and
 (4) the additive free energy ${\cal E}_{a}$ of the water molecules located at 
 ends of the tube, which come from the interaction with the outside 
 water molecules. 
 In this model, LJ potential from the nanotube wall only 
 affect the parameter ${\cal E}_{c}$. 
 The fitted parameters obtained from MD results~\cite{Hummer2001} are in 
 agreement with the directly calculated or estimated values from the 
 corresponding interactions, that is,
 the obtained equilibrium properties are well consistent with the results of 
 MD simulation for both the unmodified and modified systems. 
 In the LG model, by analyzing transition states in which one or 
 two water molecules partially occupied the end(s) of the tube, the whole free 
 energy
 landscape, and the main reaction paths can be 
 obtained, therefore, the two-state like kinetics of filling and emptying 
 of water can be clearly explained. 
 
 Considering a nanotube filled with water molecules, we first partition the 
 tube into $N$ cells along its long axis, 
 the length of every cell is corresponding to the length of hydrogen bond 
 between water molecules. We form our model step by step: (1) we consider 
 one water molecule moves into a 
 cell from the bulk water, where the cell is only a box without any
 interaction except limiting the water molecule in volume $V_{0}$. In this 
 process, the changed free energy is 
 $\Delta \mu = -{\mu}_{w} + {\mu}_{g}$,
 where ${\mu}_{w}$ is the chemical potential of bulk water, 
 ${\mu}_{g} = t \mbox{ } {\ln}({\lambda}^{3}/V_{0})$, is the chemical 
 potential of 
 ideal gas with density $1/V_{0}$, $t$ is a normalized temperature. In this 
 paper, the energy unit is set as the room 
 temperature $300$ {\rm K}, if no explicit unit is written. 
 ${\lambda} = (2 \pi {\hbar}^{2}/m k_{B} T)^{1/2} \approx 0.23$ {\rm \AA} is 
 the de Broglie thermal wavelength of water.
 Actually, $\Delta \mu$ is nothing but 
 $-{\mu}^{ex}_{nt}$, where ${\mu}^{ex}_{nt}$ is the local excess chemical 
 potential of water inside the nanotube, about $-6.87$ 
 {\rm kcal/mol} $\approx -11.6$ from MD simulation~\cite{Hummer2001}. 
 (2) if there is interaction $V_{e}({\mathbf r})$ between the wall of the 
 cell and water molecule, except $\Delta \mu$, an additive free energy is 
  \begin{eqnarray}
 {\cal E}_{c} = - t \mbox{ } {\ln}({1 \over V_{0}} 
 \int_{V_{0}} d {\mathbf r} \mbox{ } {\exp}[- V_{e}(\mathbf r)/t]).
 \label{LJenergy}
 \end{eqnarray}
 For nanotubes, $V_{e}({\mathbf r})$ is the Lennard-Jones(LJ) potential 
 between the water 
 molecule and all carbon atom of the nanotube. Calculation 
 shows that $V_{e}({\mathbf r})$ is almost independent on the azimuth $\phi$ 
 and axis position $z$ in infinite nanotubes~\cite{Tu2002}, For finite
 nanotubes, the result is also right, except a $z$ dependence 
 near the ends of the nanotubes. 
 In $(6, 6)$ nanotube, due to its very small radius, for both the unmodified 
 and modified LJ parameters, $V_{e}(r)$ 
 can be well approximated as a box potential well, whose widths 
 (about $0.8$ {\rm \AA}) are in agreement with the interior radius of the 
 tube found in 
 MD~\cite{Hummer2001}, and the depths of well are about
 $-8 \sim -9$ and $-5 \sim -6$, respectively. 
 It is not very difficult to numerically calculate the more exact value of 
 ${\cal E}_{c}$, but it is not necessary; 
 (3) if two water molecules locate at two near adjacent cells, a hydrogen
 bond will be formed, the interaction is denoted $V_{H}(\mathbf r)$. 
 Under the box potential well approximation of 
 $V_{e}(r)$, the free energy can be simply written as,
 \begin{eqnarray}
 {\cal E}_{H} = -t \mbox{ }{\ln} {1 \over V_{0}^{2}} 
 \int_{V_{1}} d {\mathbf r}_{1} \int_{V_{2}} d {\mathbf r}_{2} \mbox{ }
 {\exp}[- V_{H}({\mathbf r}_{1} - {\mathbf r}_{2})/t],
 \label{hydrogenbond}
 \end{eqnarray}
 where, due to the limit of the length of hydrogen bond and the 
 nanotube wall, 
 the two water molecules will occupy two cells $V_{1}$ and $V_{2}$ (both 
 volumes are $V_{0}$), respectively. The freedom degree ${\mathbf r}_{2}$ is 
 limited, only some rotation can be allowed, we have, 
\begin{eqnarray}
{\cal E}_{H} = E_{H} - t \mbox{ } 
 {\ln}( {a {\lambda}^{3} \over V_{0}}) 
 = E_{H} - {\mu}_{g} - {\mu}_{i},
\label{freeEh}
\end{eqnarray}
where $E_{H}$ is the energy of the hydrogen bond. All other contributions, 
including the rotation of the non-spherical water molecules
, the rotation of water molecules
 around their hydrogen bond, and the possible angle changes of the hydrogen 
 bonds, etc., are writen as the factor $a$ ($a>1$ indicates the existence of
 these freedom degrees), the corresponding free energy is noted as 
 ${\mu}_{i}$.
 While the number ($n$) of water molecules formed a continuous hydrogen-bond 
 chain inside the tube is not very large, we suppose that 
 $\mu_{i}$ is independent on $n$;
 (4) if a water molecule locate at ends of the nanotube, 
 besides ${\cal E}_{c}$ (while water is partially out of the nanotube, 
 even ${\cal E}_{c}$ also increase), there is an additive free 
 energy ${\cal E}_{a}$ from the interaction between the water molecule and 
 the outside water. In fact, ${\cal E}_{a}$ is also from hydrogen bonds of 
 water molecules, 
 but due to the lower density of bulk water, 
 it is higher than ${\cal E}_{H}$.
 
Based on these considerations, the water molecules inside the nanotube can be
treated as a simply 1D LG model with $N$ sites, including three parameters. 
Only ${\cal E}_{c}$ is dependent on LJ parameters, and the 
dependence can be approximately calculated. Hamiltonian of the model is,
\begin{eqnarray}
{\cal H} = \sum_{i=1}^{N} {\cal E}_{i} \mbox{ } c_{i}
 + {\cal E}_{H} \sum_{i=1}^{N-1} c_{i} c_{i+1}, 
 \label{hamiltonian}
\end{eqnarray}
where ${\cal E}_{i} = \Delta \mu + {\cal E}_{c} + 
(\delta_{i,1} + \delta_{i,N}) \mbox{ } {\cal E}_{a}$, 
$\delta_{i,j}$ is Kronecker delta, $c_{i} = 1$(occupied), or 
$0$(unoccupied). The model is equivalent to an Ising model, if we transform
$s_{i} = 2 c_{i} - 1$.
For a 1D Ising
model, if the external field $H=0$, the states with $M$ and $-M$ are symmetry,
 and the energy barrier is very low, 
where $M = \frac{\sum s_{i}}{N}$. In the LG model, 
 the corresponding external field 
$H= -({\cal E}_{i} + {\cal E}_{H})/2$. If ${\cal E}_{c}$ is very low, 
$H >0$, water molecules can fill the nanotube, while 
${\cal E}_{c}$ increases (the well depth decreases), $H$ decreases.
If $H \sim 0$, there is an equilibrium between the filling and 
emptying states. 
However, some details must be 
considered as we try to explain the MD results of Ref.~\cite{Hummer2001} by 
using the LG model: (1) the finite sites $N$; (2) the additive energy 
${\cal E}_{a}$ at ends. 
Since $N$ is not very large, we can directly enumerate all $2^{N}$ 
states.
For general case, Hodak and Girifalco~\cite{Hodak2001} have written the 
partition function for the LG model with any finite sites and the additive 
energy. 

\begin{figure}
\centerline{\epsfxsize=10cm \epsfbox{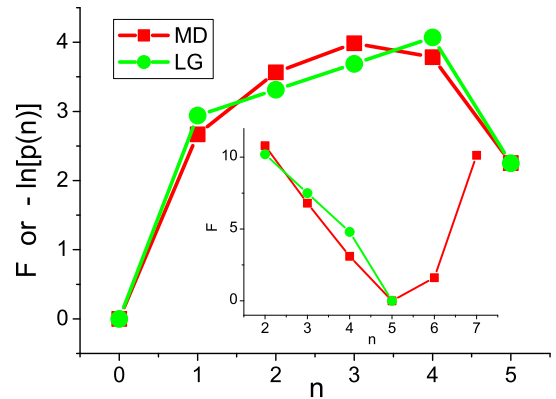}}
\caption{ For the system with modified LJ parameters (see text), the
free energy from LG model and $-{\ln} \mbox{ } p(n)$ from MD 
simulation are plotted, where $p(n)$ is the
possibility of finding exactly $n$ water molecules inside the nanotube,
the simulation data come from Ref.{\protect\cite{Hummer2001}}. The results 
of the
system with unmodified LJ parameters are shown at the inset.
\label{fig1}}
\end{figure}

For unmodified and modified LJ parameters, Hummer {\it et al.} have given the
probability of finding exactly $N$ water molecules inside the nanotube
(Figure 2 of Ref.~\cite{Hummer2001}). In their MD simulation, $N=5$. We can 
obtain
the model parameters (${\cal E}_{H} \approx -6.27$, 
${\cal E}_{a} \approx -2.93$, and ${\cal E}_{c} \approx -4.96$) by fitting the 
MD probability data of the modified system. 
The regenerated free energy profile is 
very well in agreement with the MD probability distribution in the modified 
system as shown in Fig.\ref{fig1}. Here, the fitting 
${\cal E}_{c}$ is well consistent with the estimated value
from LJ potential by Eq.(\ref{LJenergy}). The hydrogen bond of water 
inside the nanotube is about $7$ {\rm kcal/mol} $\approx 11.8$ 
(Figure 3(a) of Ref.~\cite{Hummer2001}), is stronger
than their average value in bulk water 
($5$ {\rm kcal/mol} $\approx 8.4$). From 
Eq.(\ref{freeEh}), we know that the free energy contribution ${\mu}_{i}$ of 
all rotational
freedom degrees is about $0.7$. The ${\cal E}_{a}$ is only half of 
${\cal E}_{H}$, the physical reason is the lower density of bulk water.
Since both ${\cal E}_{a}$ and ${\cal E}_{H}$
are independent of LJ potential, their values can be used in the unmodified 
system.
For the large statistic errors of MD data in the unmodified system, 
replacing to fit MD results, we directly use that 
estimated ${\cal E}_{c} \approx -8$. The obtained free energy 
profile is shown in inset of Fig.\ref{fig1}, it is
approximately in agreement with the MD probability distribution. 
For equilibrium states, due to the large ${\cal E}_{H}$, 
water is favor to the states with a single sequentially occupied 
sites, for example, $P(10100)$ is only about 
$1/500$ of $P(11000)$, where $P$ is the probability of states. Similarly, 
water is favor to the states whose end sites are occupied. For example,
$P(01100)$ is only about $1/20$ of $P(11000)$ due to ${\cal E}_{a}$. 
For a good approximation, we can calculate 
the partition function only considering
the states with a single sequence of hydrogen-bond chain 
and at least occupied an end of the tube. 
We call it as single-sequence end occupied approximation (SSEA).
In a very current MD 
simulation~\cite{Waghe2002}, for short nanotube($13.5$ {\rm \AA}) and long
nanotube($27$ {\rm \AA}), Hummer and coworkers carried to a longer time MD 
simulation, by changing LJ parameters by varying a control variable $\eta$, 
$\epsilon = {\epsilon}_{1} {\eta}^{2}$, and 
$\sigma = {\sigma}_{1}/{\eta}^{1/6}$, where $\eta = 1.0$ is 
corresponding to the unmodified LJ parameters. The model can very 
well explain these results. For example, (1) very rare non-SSEA 
events are found; (2) the approximated linear free energy profile $G(n)$ for 
$1 \le n < N$ (Using the SSEA of the LG model, 
$G(n) = n \Delta G + const$, 
where $\Delta G = \Delta \mu + {\cal E}_{c} + {\cal E}_{H}$.); 
(3) $\Delta G$ is proportional to $\eta$ (Our calculations show that the 
LJ interaction well depth is proportional to $\eta$, therefore
$\Delta G$ is proportional to $\eta$.). 
Actually, for explaining the equilibrium property found in their MD, Hummer 
and coworkers have formed 
a free energy formula (Eq.(3) of Ref.~\cite{Waghe2002}), which is 
equivalent to the SSEA of our model.

\begin{figure}
\centerline{\epsfxsize=6cm \epsfbox{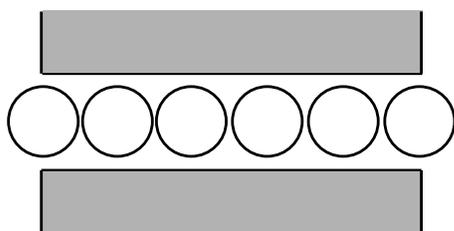}}
\caption{ $n=6$ water molecules are found inside the nanotube, 
where two water molecules only partially occupy at two ends of the nanotube.
\label{fig2}}
\end{figure}

The equilibrium property of water filled into nanotube has been very well
described in the LG model. Before investigating the kinetics, we discuss that 
the states with $6$ water molecules inside the nanotube, its probability is
not very small in MD simulation~\cite{Hummer2001,Waghe2002}. It may be 
corresponding to a state where the ends of the nanotube are partially 
occupied by two water molecules as shown in Fig.\ref{fig2}. 
The additive free energy of a molecule partially occupied is denoted as 
${\cal E}_{a}^{\prime}$, which is different from ${\cal E}_{a}$, due to both 
differences of carbon-water interaction and the interaction with the outside 
 water. Since water molecules must pass through these partially
occupied states as their filling or emptying, the free energy may be 
comparable with the kinetic energy barrier. For the states which $n < 6$, 
the effects of the partially occupied sites can be ignored, then we have 
${\cal E}_{a}^{\prime} = {\cal E}_{a} + 
[G(6)-G(5) - {\cal E}_{H} - \Delta \mu - {\cal E}_{c}]/2$.
We obtain that ${\cal E}_{a}^{\prime}$ is $-0.8$ and $-1.76$ for 
unmodified and 
modified system, respectively. We write that 
${\cal E}_{a}^{\prime} = (\alpha - 1) {\cal E}_{c} + {\delta}_{w}$, 
where $\alpha$ is a corrected factor of the carbon-water 
interaction while water molecule partially move out of the nanotube, and 
${\delta}_{w}$ is the contribution from the outside water near the end of
nanotube. 
If we suppose that both $\alpha$ and ${\delta}_{w}$ are independent of LJ 
parameters, we have $\alpha \approx 0.67$ and ${\delta}_{w} \approx -3.4$.
The obtained factor $\alpha$ is well in agreement with the calculated result 
of LJ potential while the water molecule is halfly out of the tube, 
and ${\delta}_{w}$ is comparable to ${\cal E}_{a}$. It maybe indicate a real 
physical picture.

The kinetics of filling and emptying is determined by the real reaction 
paths. The existence of two minimums of the free energy profile $G(n)$ 
is not a sufficient condition for the two-state like behavior of filling and
emptying. The understanding of the whole free energy landscape 
is necessary. 
Generally, a rate equation can be written as,
$\frac{d P_{a}}{d t} = \sum_{b} k(b \rightarrow a) P_{b} - 
\sum_{b} k(a \rightarrow b) P_{a}$,
where $P_{a}$ is the possibility of state $a$, $k(b \rightarrow a) = 
\gamma \mbox{ } {\exp}[-\beta (F_{a,b} - F_{b})]$ is the 
rate from state $b$ to $a$, 
where $\gamma$ is a factor, $\beta =1/k_{B} T$, $F_{a,b}$ is the free energy
barrier or saddle point (transition state) separating the 
state $b$ and state $a$~\cite{Zwanzig1997,Hangii1990}, $F_{b}$
is the free energy of the initial state $b$. Obviously, only two connected 
states which they have a common boundary can transit each other. 
For the filling, emptying or transporting of water, there are 
$2^{N}$ states (here we have supposed that the outside water molecules be 
always in equilibrium). The basic motion is that 
 a water molecule jump a step inside nanotube or pass through
an end of the tube. 
Since the barrier of broken hydrogen bonds is too high, 
the collective motion of a hydrogen-bond chain often 
occure~\cite{Berezhkovskii2002}, 
it means all molecules of the chains move a basic step at 
the same time. For any cases, the axis positions $z$ of the all moving 
molecules are good reaction coordinates. 
In the motion of passing an end of nanotube, the 
free energy barrier locates nearly at the end 
point of the tube (molecule half occupy the end cell), so we 
approximately
have the barrier ${\cal E}_{b} \sim {\cal E}_{a}^{\prime}$, but 
for the motion of water inside of the tube, the barrier is very small 
($\approx 0$) due to the  independence of LJ potential on $z$. 
 For the collective motion, the free energy barrier is the sum of the 
 barrier of every moving molecule.
 Besides this, the facter $\gamma$ is different from that of 
single molecule motion. Actually, $\gamma$ is dependent on the thermal 
velocity $v_{T} \sim \sqrt{ k_{B} T/m} \propto 1/\sqrt{n}$, where $n$ is 
the number of molecules in the collective moving chain. While the length of
a cell $l$ is very small, $\gamma = v_{T}/l$~\cite{Chou1998}, 
so $\gamma \approx 1.5/\sqrt{n}$ {\rm ps}$^{-1}$. 

Consequently, based on the simple estimation, we can draw the whole free 
energy landscape. Obviously, 
any non-SSEA state can fastly transit to the connected SSEA states, but the 
inverse transition is very slow, and the barrier between neighbor SSEA states 
 is lower than that between the SSEA states and non-SSEA states. 
For example, according to our estimation, from $(10000)$ to $(11000)$, the 
overcome barrier is about 
$\Delta G + {\cal E}_{a}^{\prime} - {\cal E}_{a} \approx 1.4$,
but from $(10000)$ to an non-SSEA state $(10001)$, the overcome energy is 
about $\Delta \mu + {\cal E}_{c} + {\cal E}_{a}^{\prime} \approx 4.8$.
Therefore, the reaction paths almost do not pass through 
the non-SSEA states in the filling, emptying and transporting processes.
It indicates that SSEA describes the kinetics behavior very well.
Based on their MD results that the probability of non-SSEA states is very 
low at equilibrium, Hummer and coworkers have formed 
 a kinetic model only including the reaction between the neighbor SSEA states 
 (Eq.(2) of Ref.~\cite{Waghe2002}), and found that it worked very well. 
 However, the reason of all reaction paths including 
 non-SSEA states can be neglected is the high free energy barriers, not 
 only the high free energy itself of the states.
 
Now, we can easily explain the two-state like behavior of filling and 
emptying: water molecules collectively moving a step into or out from the 
nanotube, the barriers of the 
middle states ($1 \le n <N$) are small (about $1.4$ and $1.1$, respectively), 
and the effect of the pre-exponent 
factor is not obvious, so the lifetime of these states is very short, the
 filling and emptying processes have a two-state like behavior. 
Actually, we can calculate the rate constant of every reaction and the 
lifetime of every middle state. 
For example, we have that  
$k(10000 \rightarrow 00000) \approx 0.5$, the rate
constant of the inverse process is about $0.012$, where unit is 
{\rm ps}$^{-1}$. The order of magnitude of the values is in agreement with 
the estimated results of MD (about $0.3$ and $0.009$~\cite{Waghe2002},
respectively.).  

As this work was being finished, we noticed a work~\cite{Maibaum2002}, 
in which Maibaum and Chandler treated whole system 
consisting of the channel of nanotube and the outside bulk liquid as a 
lattice gas by a MC simulation. It was shown that there are similar behavior 
found in MD simulation in generic liquid states. Under some approximations, 
they also analytical analyzed the equilibrium property of the system. 
However, for quantitatively explaining the behavior of water confined in 
nanotube, we should consider the characters of water, for example, 
the average density of bulk water is only $1/5$ of that of the filled state 
in the tube, then the 
interaction (${\cal E}_{a}$) between a water molecule at an end of the tube
 and the outside 
molecules is weaker than the interaction between two adjacent 
inside molecules. 
For the equilibrium property of system, 
the whole outside bulk water can be thought as an aqueous reservoir and 
contribute an 
additive free energy ${\cal E}_{a}$ to the inside molecules at ends of 
nanotube, but
for the kinetics, we should identify  
these states, in which different number outside water molecules contact
 with an inside molecule one by one by hydrogen bonds.
 If we denoted the state in which $m$ outside molecules contact with $n$ 
 inside molecules one by one as ``$(m){\_}[n]$''~\cite{Zhou1}, then from 
 the state, after a filling or emptying step, the final state is 
 ``$(m-1){\_}[n+1]$'' or ``$(m+1){\_}[n-1]$'', 
 respectively. The complete rate chain will consist of all states with 
 any $m$ (from $0$ to $\infty$). However, since the longer outside molecule 
 chain is more instable, in our interesting time scale, the 
longer chain will be thought as rapidly equilibrating with the surrounding 
water. Depending on the considered largest $m$, we have
different order approximations of the complete rate equation. 
In this paper, we used the lowest order ($0-$order) approximation in which
 we supposed that 
one outside molecule rapidly contacted with and departed from the end inside 
molecule, the equilibrium can rapidly reach in our interested time scale. 
If we consider higher order ($1-$order) approximation, 
the details of energy barrier will be changed, but 
the differences between the two approximations is not very large, so we think 
that the principal character
of the kinetics of the system has been captured by the simple 1D LG model. 
We wish to give more quantitative agreement on rate constants by using higher
order approximation in next work.
 

X. Z. thanks Professor Hummer for sending him their MD data and a new 
paper of Ref.~\cite{Waghe2002} and telling him the work of Maibaum and 
Chandler (Ref.~\cite{Maibaum2002}). X. Z also thanks Professor Chandler 
for comments on an earlier draft of this paper, and Mr. Z.-C. Tu for 
discussions. X. Z is financially supported by the Grants-in-Aid for 
Scientific Research of JSPS. 

(Correspondent address: zhou@pe.titech.ac.jp).

\begin{thebibliography}{99}
\bibitem{Zhou1} The right contacted states are similar as the left contacted
states, denoted as ``$[n]{\_}(m)$''. For filled state, it is possible to be 
contacted by outside molecules in two directions, denoted as 
``$(m_{1}){\_}[N]{\_}(m_{2})$''.
\end{thebibliography}
\end{document}